\documentclass[12pt,final]{amsart}
\usepackage{amsmath, amssymb}

\usepackage{graphicx,setspace}

\usepackage{amsfonts}

\def\calf{\mathcal{F}}
\def\E{\mathcal{E}}

\def\R{\mathbb{R}}

\def\A{\mathcal{A}}

\def\L{\mathcal{L}}
\def\U{\mathcal{U}}
\def\n{^{(n)}}
\def\I{\mathbb{I}}

\numberwithin{equation}{subsection}
\newtheorem{thm}{Theorem}[subsection]
\newtheorem{lem}[thm]{Lemma}

\newtheorem{cor}[thm]{Corollary}
\newtheorem{prop}[thm]{Proposition}
\newtheorem{RM}[thm]{Remark}

\usepackage[comma, authoryear]{natbib}

\usepackage[color, notref, notcite]{showkeys}
\usepackage{enumerate}
\definecolor{labelkey}{rgb}{0.6,0,1}

\begin{document}

\title{Optimal hedging in discrete time}

\author{Bruno R\'emillard}

\address{Service de l'enseignement des m\'ethodes quantitatives de
gestion and GERAD, HEC Montr\'eal, 3000 chemin de la C\^{o}te
Sainte-Catherine, Montr\'eal (Qu\'ebec) Canada H3T 2A7}
\email{bruno.remillard@hec.ca}

\author{Sylvain Rubenthaler}
\address{Laboratoire J.-A. Dieudonn\'{e}, Universit\'{e} de Nice-Sophia
Antipolis, Parc Valrose, 06108 Nice Cedex 2, France}
\email{rubentha@unice.fr}

\thanks{Partial funding in support of this work
was provided by the Natural Sciences and Engineering Research
Council of Canada, by the Fonds Qu\'{e}b\'{e}cois de Recherche sur la Nature
et les Technologies, the Institut de Finance Math\'ematique de
Montr\'eal and by the PPF Complexit\'e-Mod\'elisation-Finance de
l'Universtit\'e Nice-Sophia Antipolis. We would like to thank Gerasimos Rassias for his helpful suggestions.}

\maketitle

\begin{abstract}
Building on the work
of \citet{Schweizer:1995a} and \citet{Cerny/Kallsen:2007}, we present discrete time
formulas minimizing the mean square hedging error for multidimensional assets.
In particular, we give explicit formulas when a regime-switching random walk or
a GARCH-type process is utilized to model the returns. Monte Carlo simulations
are used to compare the optimal and delta hedging methods.
\end{abstract}





\section{Introduction}
\setcounter{equation}{0}

In many applications, one is interested in finding a discrete-time dynamically traded portfolio so that its value at maturity is as close as possible to a target function of the
underlying assets. When the target function is a payoff, this can be interpreted
as option pricing and hedging. However, sometimes the target function is not a
payoff as it happens when one tries to replicate hedge funds or create synthetic
funds \citep{Papageorgiou/Remillard/Hocquard:2008}. Assuming that the error measure
is the average quadratic hedging error, \citet{Schweizer:1995a} solved the hedging
problem for one risky asset.
He showed
 that the initial value of the portfolio, which can be interpreted as the ``value'' of the option,
is the average, under the ``real probability measure'', of the
discounted payoff, multiplied by a martingale, which is not
necessarily positive. In the latter case,  the martingale cannot be
used as the density of an equivalent martingale measure.

Even if the hedging problem has been solved quite generally by
\citet{Schweizer:1995a} in the one-dimensional case, it seems to
have been ignored or forgotten, e.g., \citep{Bouchaud/Potters:1999}
or \citet{Cornalba/Bouchaud/Potters:2002}. More troubling, delta
hedging, based on the Black-Scholes-Merton model, is still used in
practice even if it has been shown  that the geometric Brownian motion
model is an inadequate model for the underlying assets
\citep{Kat/Palaro:2005}. Even when the geometric Brownian motion
model is adequate, the hedging error in discrete time is not zero,
converging only to zero as the number of hedging periods tends to
infinity. See, e.g., \citet{Boyle/Emanuel:1980},
\citet[Chapters 46-47]{Wilmott:2006c}. In addition, when the market
is not complete but there is no arbitrage, there are infinitely many
equivalent martingale measures (EMM). One then has to choose the
``best'' martingale measure with respect to some utility criterion.
There is a huge literature on this subject. Indeed, when a NGARCH process is utilized to model the log-returns,
\citet{Duan:1995} proposes a solution to the EMM problem. Unfortunately, Duan also suggests a delta hedging strategy, which has been shown to be wrong by
\citet{Garcia/Renault:1998}.

Motivated by applications in hedge fund replication,
\citet{Papageorgiou/Remillard/Hocquard:2008} proposed  a locally
optimal solution minimizing the average quadratic hedging error at
each period for the general multidimensional asset case. They
erroneously claimed that it was globally optimal, which is only true
if the discounted underlying assets are martingales. A first
motivation for the present paper is to correct that mistake and give
explicit formulas for the results in \citet{Cerny/Kallsen:2007} and
generalizing those of \citet{Schweizer:1995a}. A second motivation is to show that,
when regime-switching random walks and GARCH processes are used to model the returns,
the optimal solution of the hedging problem yields superior results to those obtained by delta hedging.

The optimal solution of the discrete time hedging problem is
described in Section \ref{sec:discret}, giving explicit expressions
for the results of \citet{Cerny/Kallsen:2007}. It is worth noting
that when the asset value process is Markovian,  or a component of a
Markov process, the optimal solution can be implemented using
approximation techniques of dynamic programming. Two such cases are
considered. Finally, in Section \ref{sec:application}, simulations
are used to compare optimal hedging with delta hedging for geometric
random walks and NGARCH models.

\section{Optimal hedging strategy in discrete time}\label{sec:discret}

Denote the price process by $S$, i.e., $S_k$ is the value of the $d$ underlying assets at period $k$ and let $\mathbb{F} =
\{\calf_k, k=0,\ldots, n\}$ be a filtration under which $S$ is adapted. Assume that $S$ is square integrable.
Set $\Delta_k = \beta_k S_k-\beta_{k-1}S_{k-1}$, where the discounting factors $\beta_k$ are predictable, i.e. $\beta_k$
is $\calf_{k-1}$-measurable  for $k=1,\ldots, n$.

The aim of this section is to find an initial investment amount
$V_0$ and a predictable investment strategy $\overrightarrow{\phi} =
(\phi_k)_{k=1}^n$ such that $\phi_k^\top \Delta_k$ is square
integrable and which minimizes the expected quadratic hedging error
$E\left[ \left\{G\left(V_0,\overrightarrow{\phi}\right)
\right\}^2\right]$, where $ G =
G\left(V_0,\overrightarrow{\phi}\right) = \beta_n C - V_n$, and $
V_k = V_0+\sum_{j=1}^k \phi_j^\top \Delta_j$, $k=0,\ldots,n$.

The existence of an optimal solution was proven first in the
univariate case by \citet{Schweizer:1995a}. \citet{Motoczynski:2000}
considered a multivariate setting without furnishing explicit
solutions. Finally, \citet{Cerny/Kallsen:2007} treated a much more
general case. However, in the discrete time case, it is faster and
easier to find directly the explicit formulas than trying to recover
them from their results.

\subsection{Offline computations}

Once a dynamic model is chosen for the asset prices, one must start
with some  computations that are necessary for the implementation.
 Set $P_{n+1}=1$, $\gamma_{n+1}=1$, and for $k=n,\ldots, 1$,
define $ A_k = E\left( \Delta_k \Delta_k^\top
P_{k+1}|\calf_{k-1}\right)$, $\mu_k = E\left( \Delta_k P_{k+1}
|\calf_{k-1}\right)$, $b_k = A_k^{-1}\mu_k$, $P_k =\prod_{j=k}^n
\left(1-b_j^\top \Delta_j \right)$, and $\gamma_k =
E(P_k|\calf_{k-1})$, provided these expressions exist. Under some
extra assumptions given below, it can be shown that they are indeed
well defined. The proof is given in Appendix \ref{app:lem}.

\begin{lem}\label{lem1} Suppose that $E(\gamma_{k+1}|\calf_{k-1}) A_k - \mu_k\mu_k^\top$ is
invertible P-a.s., for every $k=n,\ldots, 1$. Then $\gamma_{k} \in (0,1]$ and $A_k$ is
invertible for all $k=1,\ldots, n$. In addition
$(\gamma_{k+1})_{k=0}^n$ is a positive submartingale.
\end{lem}

\begin{RM}
In the univariate case, \citet{Schweizer:1995a} states sufficient
conditions for the validity of the assumptions of Lemma \ref{lem1}.
It is not obvious how they could be generalized to the multivariate
case. Therefore, in most applications, one has to verify these
conditions, often using brute force calculations.
\end{RM}

\subsection{Optimal solution of the hedging problem}

\begin{thm}\label{thm1} Under the assumptions of Lemma \ref{lem1},
the solution $\left(V_0,\overrightarrow{\phi}\right)$ of the
minimization problem is $V_0 = E(\beta_n CP_1)/\gamma_1$, and $
\phi_k = \alpha_k-V_{k-1}b_k$, where $ \alpha_k =  A_k^{-1} E\left(
\beta_n C\Delta_k P_{k+1} |\calf_{k-1}\right)$, $k=n,\ldots, 1$.
\end{thm}
The proof is given in Appendix \ref{app:thm}.

\subsubsection{Option value}

Let $C_k$ be the optimal investment at period $k$,  so that the
value of the portfolio at period $n$ is as close as possible to $C$,
in terms of mean square error. One could then interpret $C_k$ as the
value of the option at period $k$, for any $k =0,\ldots,n$. It then
follows from Theorem \ref{thm1} that
 $C_k$ is given by
\begin{equation} \label{eq:ck0}
\beta_k C_k = \frac{E(\beta_n
CP_{k+1}|\calf_k)}{E(P_{k+1}|\calf_k)}, \qquad k=0,\ldots,n,
\end{equation}
so one can write
\begin{equation} \label{eq:ck1}
\beta_{k-1} C_{k-1} =\frac{1}{\gamma_k}E\left\{ \beta_k C_k \left(1-
b_k^\top \Delta_k\right)\gamma_{k+1}|\calf_{k-1}\right)
=E\left(\beta_n C \U_k\cdots \U_n|\calf_{k-1}\right),
\end{equation}
where $\U_k = \frac{ E(P_{k}|\calf_k)}{E(P_{k}|\calf_{k-1})}$,
$k=1,\ldots, n+1$, while an alternative expression for $\alpha_k$ is
\begin{equation}\label{eq:alphak}
\alpha_k = A_k^{-1} E\left(   \beta_k C_k \Delta_k \gamma_{k+1}
|\calf_{k-1}\right).
\end{equation}

\begin{RM}
 Setting $\mathcal{Z}_0=1$ and $\mathcal{Z}_k = \prod_{j=1}^k \U_j$, $k=1,\ldots,
n$, one obtains that $(\mathcal{Z}_k,\beta_k C_k \mathcal{Z}_k,
\beta_k S_k \mathcal{Z}_k)_{k=0}^n$ are martingales. However, in
most applications, $\mathcal{Z}$ does not define a change of measure
unless it takes only positive values.
\end{RM}


\subsubsection{Implementation issues}

If the process $S$ is Markov and $C_n = C_n\left(S_n\right)$, then
$C_k = C_k(S_k)$, $\alpha_k = \alpha_k(S_{k-1})$, and $b_k =
b_k(S_{k-1})$. It follows that all these functions can be
approximated using the methodology developed in
\citet{Papageorgiou/Remillard/Hocquard:2008}. Another interesting
case is when $S_k$ is not a Markov process but $(S_k, h_k)$ is, even
if $h_k$ is not observable, as in GARCH and regime-switching models.
In this case, if $C_n = C_n\left(S_n\right)$, then $C_k = C_k(S_k,
h_k)$, $\alpha_k = \alpha_k(S_{k-1},h_{k-1})$, $b_k =
b_k(S_{k-1},h_{k-1})$, and . $\gamma_k = \gamma_k(S_{k-1},h_{k-1})$,
for $k=1, \ldots, n+1$.
All these functions can be approximated using the methodology
developed in \citet{Remillard/Hocquard/Papageorgiou:2010} for the
regime-switching case. Implementation of the hedging strategy then
requires predicting $h_t$.

\begin{RM}\label{rem:latent}
One could suggest to use the smallest filtration to get rid of the
unobservable process $h$ but in this case, all conditional
expectations based on $\calf_k$ would depend on all past values
$S_0,\ldots, S_k$, making it impossible to implement in practice.
\end{RM}

\subsection{Verification of the assumptions of Lemma
\ref{lem1}}\label{ssec:verif}

In what follows, we consider some interesting models used in
practice, for which it is possible to show that the assumptions of
Lemma \ref{lem1} hold true and that the optimal solution can be
computed via a dynamic program.
\subsubsection{Regime-switching geometric random walks}\label{sssec:HMM}

An interesting model, which includes geometric random walk models,
is to consider a regime-switching geometric random walk. Theses
models can display serial dependence in the log-returns and may
account for changing volatility over time. For implementation
issues, including estimation and goodness-of-fit tests, see, e.g.,
\citet{Remillard/Hocquard/Papageorgiou:2010}.

To define the process, suppose that $\tau$ is a finite homogeneous
Markov chain with transition matrix $Q$ with values in
$\{1,\ldots,l\}$ representing  the non-observable regimes and set $
\beta_k S_k ^i = S_0^i \prod_{t=1}^k \{1+ \xi_t^i\}$, $i=1,\ldots,
d$, where, given $\tau_1=i_1,\ldots,\tau_n=i_n$,
$\xi_1,\ldots,\xi_n$ are independent with $\xi_j \sim
\mathbb{P}_{i_j}$, $j=1,\ldots,n$, $\mathbb{E}_i (\xi_j) =
E(\xi_j|\tau_j=i) = m(i)$, and $\mathbb{E}_i\left(\xi_j
\xi_j^\top\right) = B(i)$. The interpretation of the model is easy:
At a given period $t$, a regime $\tau_t$ is chosen at random,
according to the Markov chain model, and given $\tau_t=i$, $\xi_t$
is chosen at random according to distribution $\mathbb{P}_i$. When
there is only one regime, one obtains a geometric random walk where
all $\xi$s are independent.

We assume that the $B(i)-m(i)m(i)^\top$ is invertible for any
$i=1,\ldots, l$. Setting $X_k = \beta_k S_k$, one gets $ \Delta_k =
X_k-X_{k-1} = D(X_{k-1})\xi_k$, $k=1,\ldots, n$, where $D(s)$ is the
diagonal matrix constructed from vector $s$. Note that $S$ is not a
Markov process in general but $(S,\tau)$ is a Markov process. The
validity of the assumptions of Lemma \ref{lem1} follows from the
next result, proved in Appendix \ref{app:res2}.

\begin{prop}\label{prop:ck}
For any $k=1,\ldots, n$ and $i=1,\ldots, l$,  $\gamma_{k}=\gamma_k
(\tau_{k-1})$, $\gamma_k(i)\in (0,1]$, $A_k =
A_k(S_{k-1},\tau_{k-1})$ and $b_k = b_k(S_{k-1},\tau_{k-1})$,  where
\begin{equation}
A_k(s,i) = \beta_{k-1}^2 D(s) \left\{ \sum_{j=1}^l
Q_{ij}\gamma_{k+1}(j) B(j)\right\} D(s),\label{eq:AHMM}
\end{equation}
\begin{equation}
b_k(s,i) = D^{-1}(\beta_{k-1}s)\rho_{k+1}(i),\label{eq:bHMM}
\end{equation}
\begin{equation}
 \gamma_k(i)=
\sum_{j=1}^l Q_{ij} \gamma_{k+1}(j) \left\{1- \rho_{k+1}(i)^\top
m(j)\right\}, \label{eq:gammaHMM}
\end{equation}
with $ \rho_{k+1}(i) =  \left\{ \sum_{j=1}^l Q_{ij}\gamma_{k+1}(j)
 B(j)\right\}^{-1} \left\{ \sum_{j=1}^l Q_{ij}\gamma_{k+1}(j)
m(j)\right\}$.
\end{prop}

%
If in addition
$C=C(S_n)$, then $C_k= C_k(S_k,\tau_k)$ and
$\alpha_k=\alpha_k(S_{k-1},\tau_{k-1})$, where
\begin{eqnarray*}
C_{k-1}(s,i) &=& \frac{\beta_k}{\beta_{k-1}} \sum_{j=1}^l
Q_{ij}\frac{\gamma_{k+1}(j)}{\gamma_k(i)}\\
 && \qquad \times \int
C_k\left\{\frac{\beta_{k-1}}{\beta_k}D(s)(\mathbf{1}+y),j\right\}
\left\{1- \rho_{k+1}(i)^\top y \right\}\mathbb{P}_j(dy),
\end{eqnarray*}
\begin{eqnarray*}
\alpha_{k}(s,i) &= &\frac{\beta_k}{\beta_{k-1}} D^{-1}(s)\left\{
\sum_{j=1}^l Q_{ij}\gamma_{k+1}(j) B(j)\right\}^{-1}\sum_{j=1}^l
Q_{ij}
\gamma_{k+1}(j) \\
&& \qquad \times \int
C_k\left\{\frac{\beta_{k-1}}{\beta_k}D(s)(\mathbf{1}+y),j\right\}y
\mathbb{P}_j(dy).
\end{eqnarray*}

\subsubsection{GARCH-type models}\label{sssec:GARCH}

Here, one assumes that $ \Delta_k  = \beta_k S_k - \beta_{k-}S_{k-1}
= \beta_{k-1}S_{k-1}\xi_k$, with $ \xi_k =
\pi_1(h_{k-1},\epsilon_k)$, and $h_k = \pi_2(h_{k-1},\epsilon_k)$
with $\pi_2$ having values in some set $\mathcal{H}$,  and where the
innovations $\epsilon_k$ are independent and identically distributed
with probability law $\nu$. It is immediate that $(S_k,h_k)$ is a
Markov process. Furthermore, almost all known GARCH(1,1) models can
be written in that way.

Suppose that for every given possible $h\in\mathcal{H}$,
$\pi_1(h,y)$ is not constant $\nu$-a.s. Using Proposition
\ref{prop:inv} and reverse induction, as in the proof of Proposition
\ref{prop:ck}, it is easy to show that the assumptions of Lemma
\ref{lem1} are met, and that for all $k=n,\ldots, 1$, $\gamma_k =
\gamma_k(h_{k-1})$ and $ A_k (s,h) =\beta_{k-1}^2 s^2 B_k(h)$, $b_k
(s,h) = \frac{m_k(h)}{s \beta_{k-1} B_k(h)}$, where $B_k(h) = \int
\pi_1^2(h,y)\gamma_{k+1}\left\{\pi_2(h,y)\right\}\nu(dy)$, $ m_k(h)
= \int \pi_1(h,y)\gamma_{k+1}\left\{\pi_2(h,y)\right\}\nu(dy)$, and
$\gamma_k(h) = \int  \left\{1 - \frac{m_k(h)}{B_k(h)}\pi_1(h,y)
\right\}\gamma_{k+1}\left\{\pi_2(h,y)\right\}\nu(dy)$. Also, if
 $C = C_n(S_n)$, then
\begin{eqnarray*}
C_{k-1}(s,h) &=& \frac{\beta_1}{\gamma_k(h)}\int C_k\left[\frac{s}{\beta_1}\{1+\pi_1(h,y)\},\pi_2(h,y)\right]\\
&& \qquad \times \gamma_{k+1}\left\{\pi_2(h,y)\right\} \left\{ 1-
\frac{m_k(h)}{B_k(h)} \pi_1(h,y) \right\} \nu(dy), \end{eqnarray*}
\begin{eqnarray*}
\alpha_{k}(s,h) &=& \frac{\beta_1}{s B_k(h)}\int C_k\left[\frac{s}{\beta_1}\{1+\pi_1(h,y)\},\pi_2(h,y)\right]\\
&& \qquad \times \gamma_{k+1}\left\{\pi_2(h,y)\right\}\pi_1(h,y)
\nu(dy).
\end{eqnarray*}

Hence, the optimal solution can be written as a dynamic program.

\section{Examples of application}\label{sec:application}

In this section we consider pricing and hedging of European calls
for two geometric random walk models, when the returns are i.i.d.
Gaussian and i.i.d. differences of Laplace distributions, and for a
NGARCH model. It follows from the previous sections that optimal
hedging solutions exist for these cases, and the optimal solution
can be written as a dynamic program associated with functions of a
finite number of variables. For solving such dynamic programs, we
discretize the state space into a finite grid and we compute
approximations of expectations using Monte Carlo simulations at
every point of the grid. Linear interpolations are used for points
outside the grid at each time step. Since the expectations are
always with respect to the same probability measure, only one
sequence of random numbers may be used, using the ideas in
\citet{DelMoral/Remillard/Rubenthaler:2006,DelMoral/Remillard/Rubenthaler:2012}.

\subsection{Geometric random walk models}

Here we consider discretized versions of the Black-Scholes (BS) and
Variance Gamma (VG) models for the underlying asset over 23 periods.
In each case, the $22$ periodic returns are i.i.d., so that the mean
and volatility at maturity are respectively $9\%$ and $6\%$. For the
BS model, the returns are Gaussian, while for the VG model, the
returns are differences of i.i.d. Gamma variates, so that the
distribution at maturity is Laplace (double exponential). These
models are particular cases of regime-switching models with only one
regime. We do not consider regime-switching models since it has been
done in \citet{Remillard/Hocquard/Papageorgiou:2010}, where the
daily log-returns of the S\&P 500 are analyzed.

\begin{figure}[h!]
 \begin{center}
\includegraphics[scale=0.375]{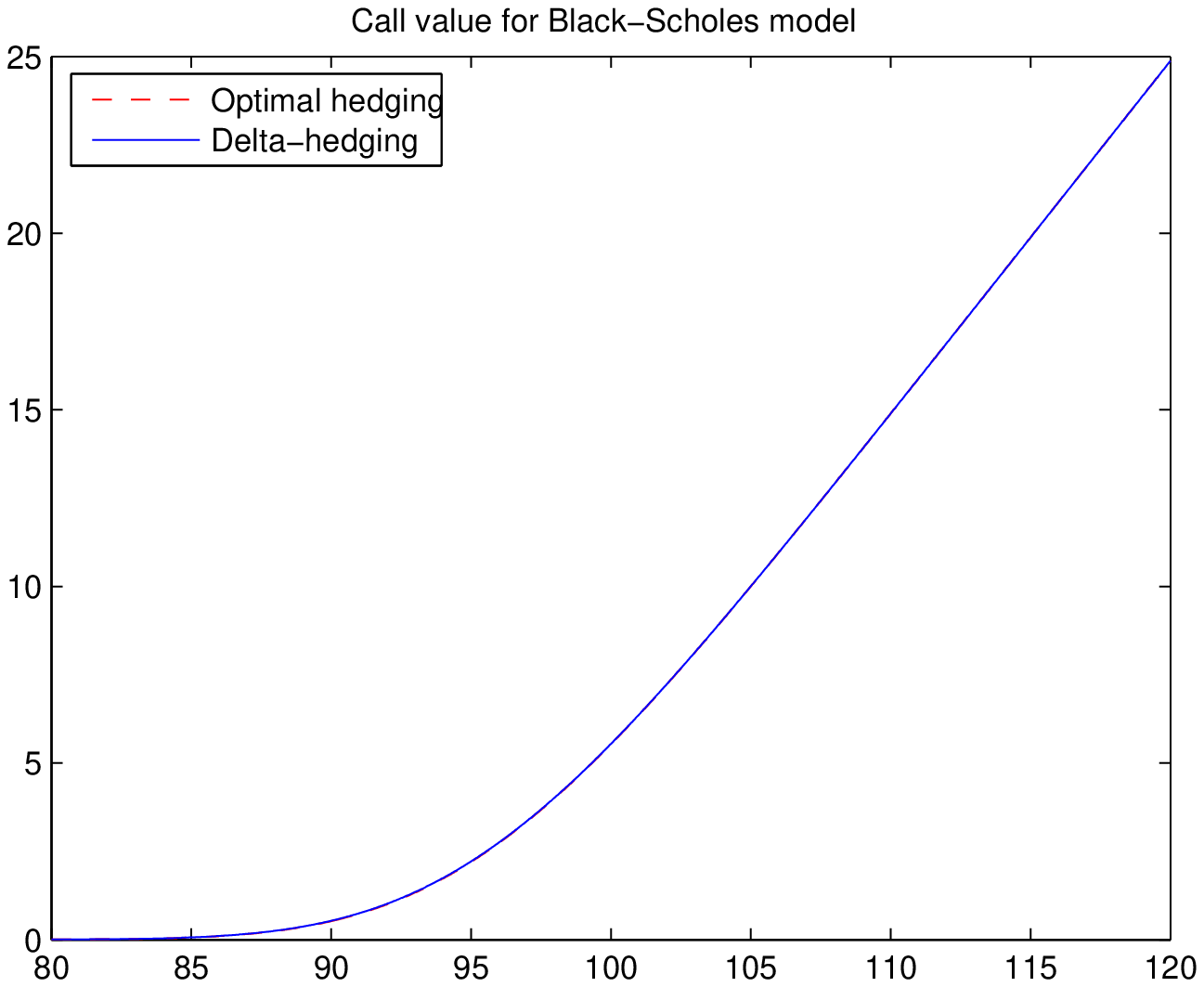}
\includegraphics[scale=0.375]{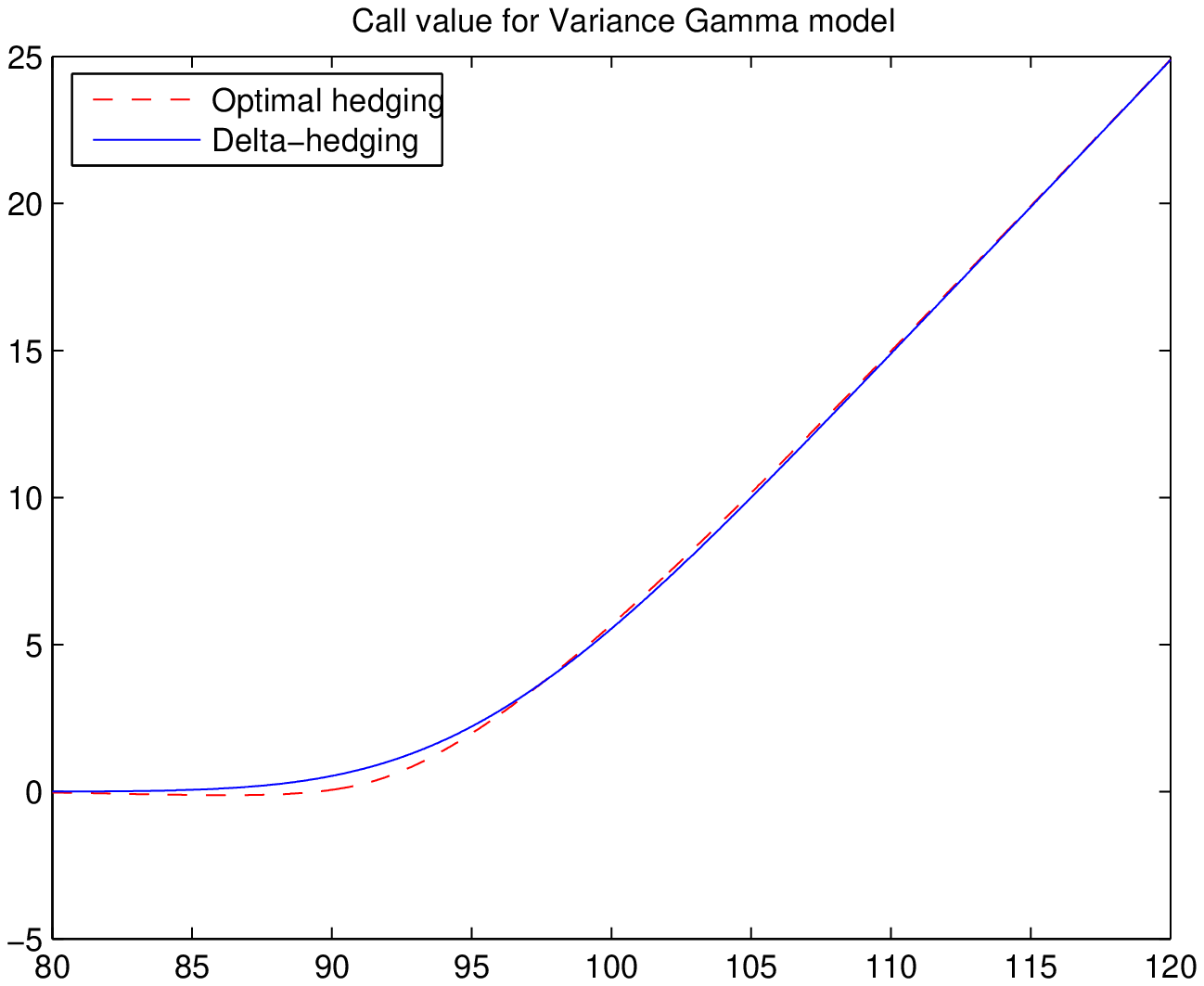}
\end{center}
\caption{Call option value $C_0$ for the Black-Scholes (left panel)
and Variance Gamma models (right panel) with $22$ periods of
hedging.}\label{fig:BSCall}
\end{figure}

\begin{figure}[h!]
 \begin{center}
\includegraphics[scale=0.375]{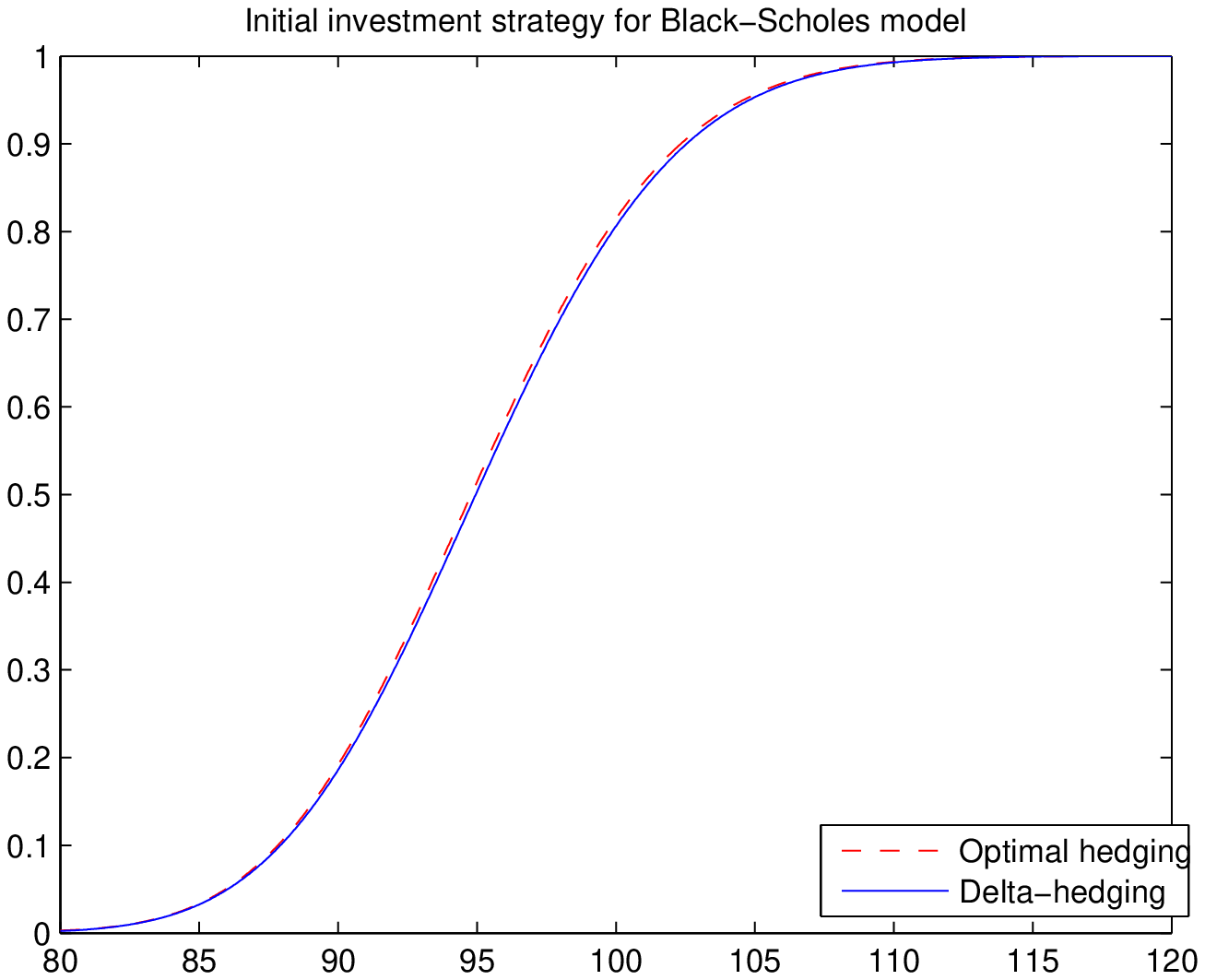}
\includegraphics[scale=0.375]{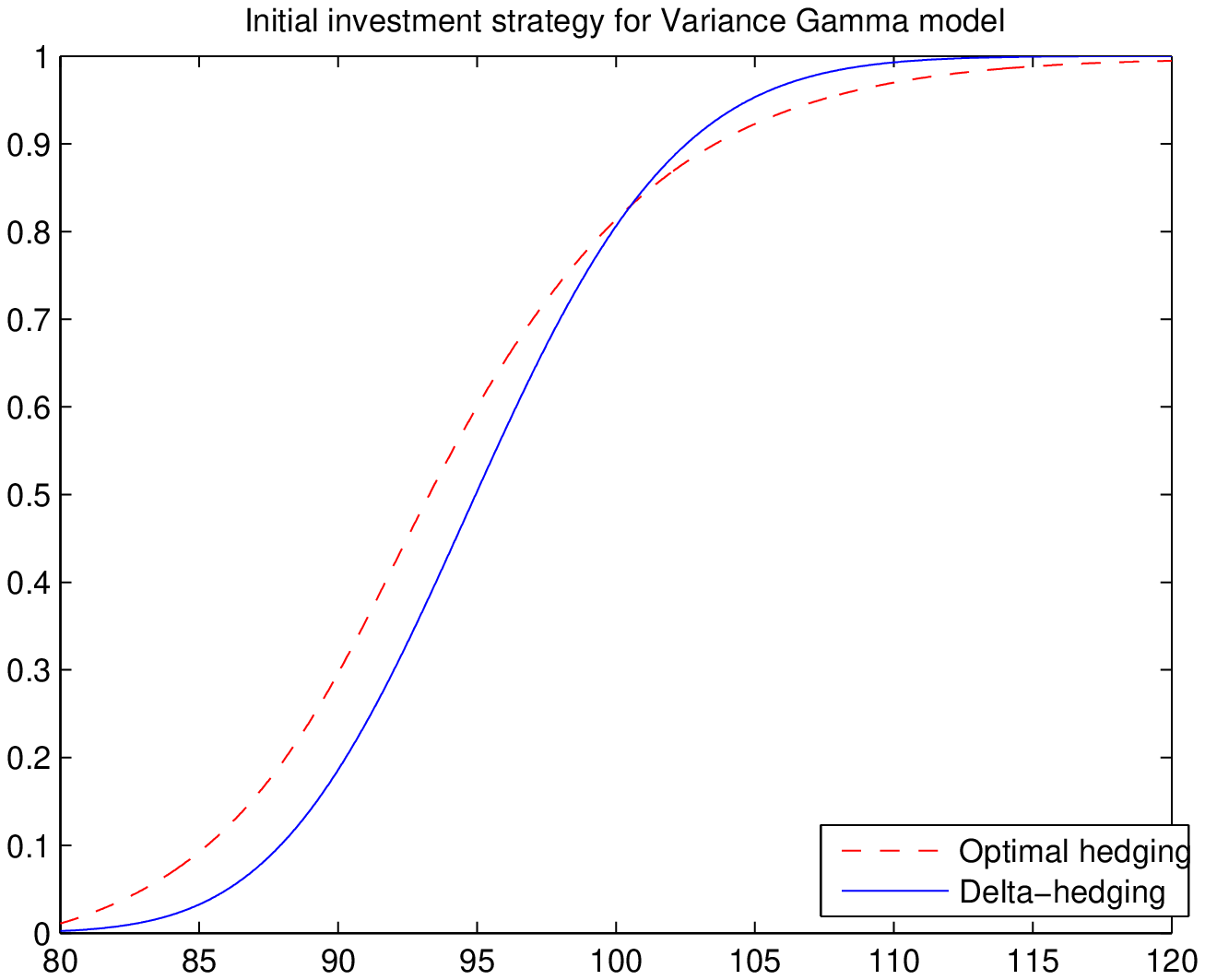}
\end{center}
\caption{Initial investment strategy $\phi_1$ in the underlying asset
for the Black-Scholes (left panel) and Variance Gamma models (right
panel) with $22$ periods of hedging.}\label{fig:BSPhi}
\end{figure}

\begin{figure}[h!]
 \begin{center}
\includegraphics[scale=0.375]{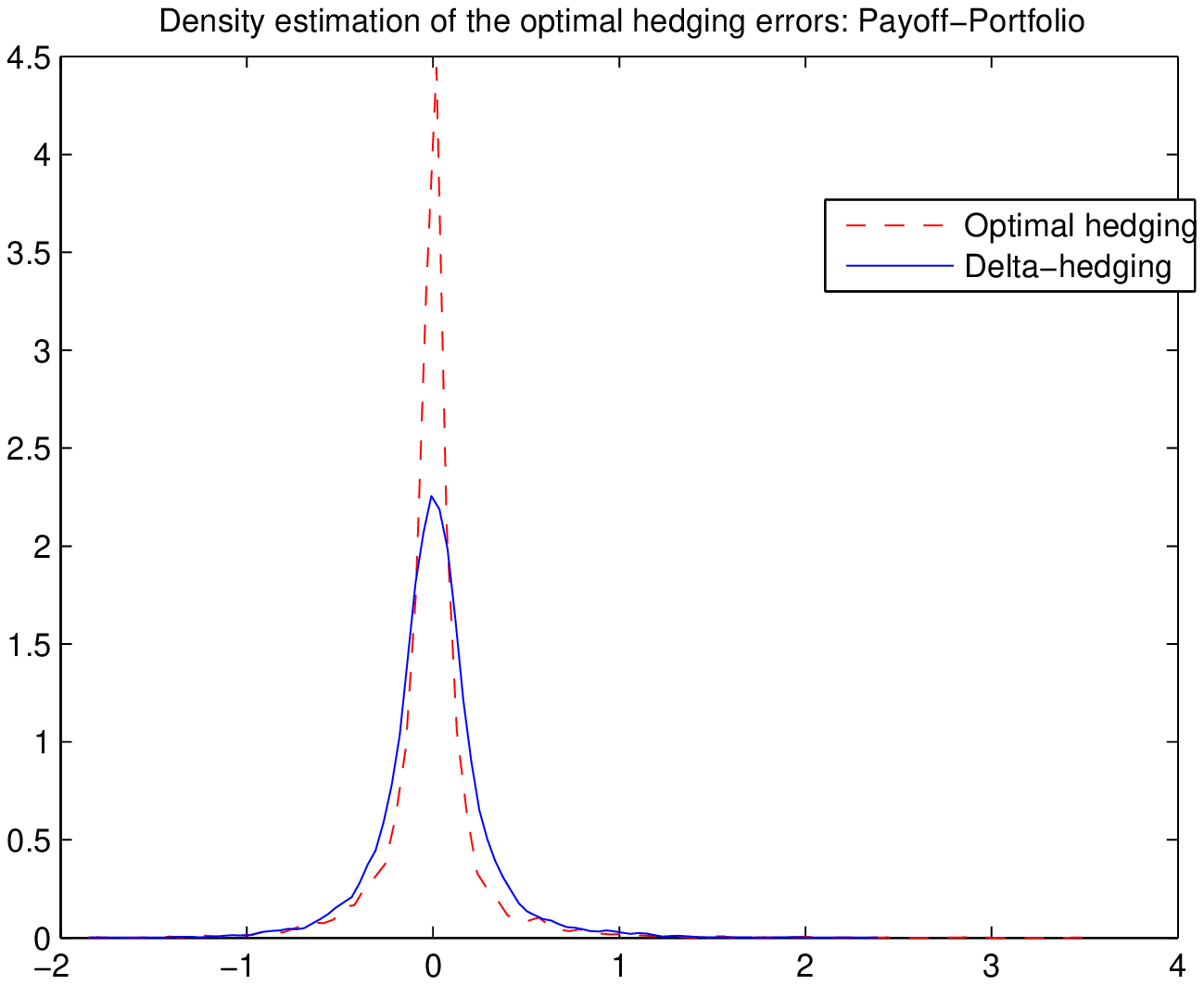}
\includegraphics[scale=0.375]{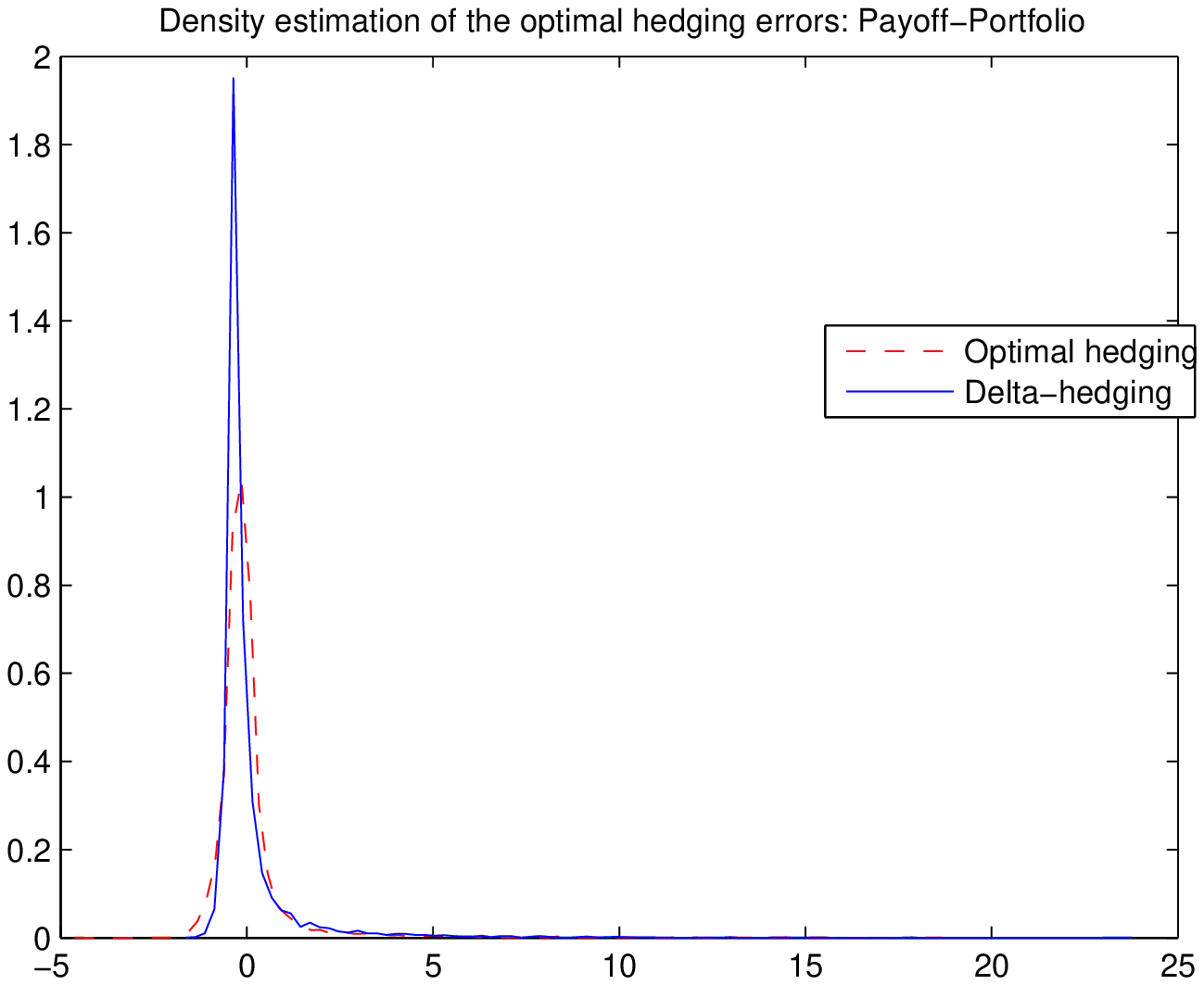}
\end{center}
\caption{Estimated densities of the hedging error $G$ for the
Black-Scholes (left panel) and Variance Gamma models (right panel)
with $22$ periods of hedging.}\label{fig:BSHE}
\end{figure}

 We are going to
price and hedge a call with strike $K=100$ and maturity $1$ year,
using 22 replication periods and a 2000 points discretization of the
asset values over the interval $[80,120]$. The annual rate is $5\%$.
A sequence of $10000$ random points were used for the computation of
functions $\alpha_k$ and $A_k$, while  $10000$ paths were used to
compute the hedging errors. Delta hedging is optimal in the
continuous time limit for the BS model, but not for the VG model. As
expected, according to Figures \ref{fig:BSCall}--\ref{fig:BSPhi},
the values of the call $C_0$  and initial investment strategy $\phi_1$,
obtained from the optimal hedging, are close to those obtained using
the Black-Scholes formula (even with 22 hedging periods), while they
differ for the VG model. This is also reflected in the distribution
of the hedging errors, as illustrated in Figure \ref{fig:BSHE}.

Descriptive statistics of the hedging errors are given in Table
\ref{tab:BS}. Simulations can also be used to show that as the
number of hedging periods increases, the hedging error tends to zero
for the BS model, while it is never $0$ for the VG model. Note that
the RMSE of the optimal hedging is always less than the one of the
delta hedging.

\begin{table}[h!]
{\footnotesize
 \caption{Statistics of hedging errors
(Payoff-Portfolio) for the Black-Scholes and Variance Gamma
models.}\label{tab:BS}
 \begin{center}
\begin{tabular}{l|r|r|r|r}
  & \multicolumn{2}{c}{Black-Scoles}   & \multicolumn{2}{|c}{Variance Gamma}\\
Stats      & Optimal        & Delta     & Optimal        & Delta   \\
\hline\hline
Average    &  -0.0065 &  0.0076  &     0.0151    &   0.0518       \\
Median     &  -0.0014 &  0.0029  &    -0.1543    &  -0.2886     \\
Volatility &   0.2537 &  0.2774  &     1.0510    &   1.2529      \\
Skewness   &   1.3781 &  0.5515  &     6.1479    &   6.0965    \\
Kurtosis   &  22.0975 &  8.8391  &    63.3897    &  60.9747     \\
Minimum    & -1.7986  & -1.7578  &    -4.4372    &  -1.5160      \\
Maximum    &  3.5389  &  2.2978  &    18.5163    &  23.6422       \\
VaR(99\%)  &  0.8231  &  0.9244  &     4.6976    &   5.9450    \\
VaR(99.9\%)&  1.9576  &  1.5779  &    11.7905    &  12.2222     \\
RMSE       &  0.2538  &  0.2775  &     1.0511    &   1.2540     \\
 \end{tabular}
  \end{center}
  }
\end{table}

\subsection{NGARCH model}

\begin{figure}[h!]
 \begin{center}
\includegraphics[scale=0.375]{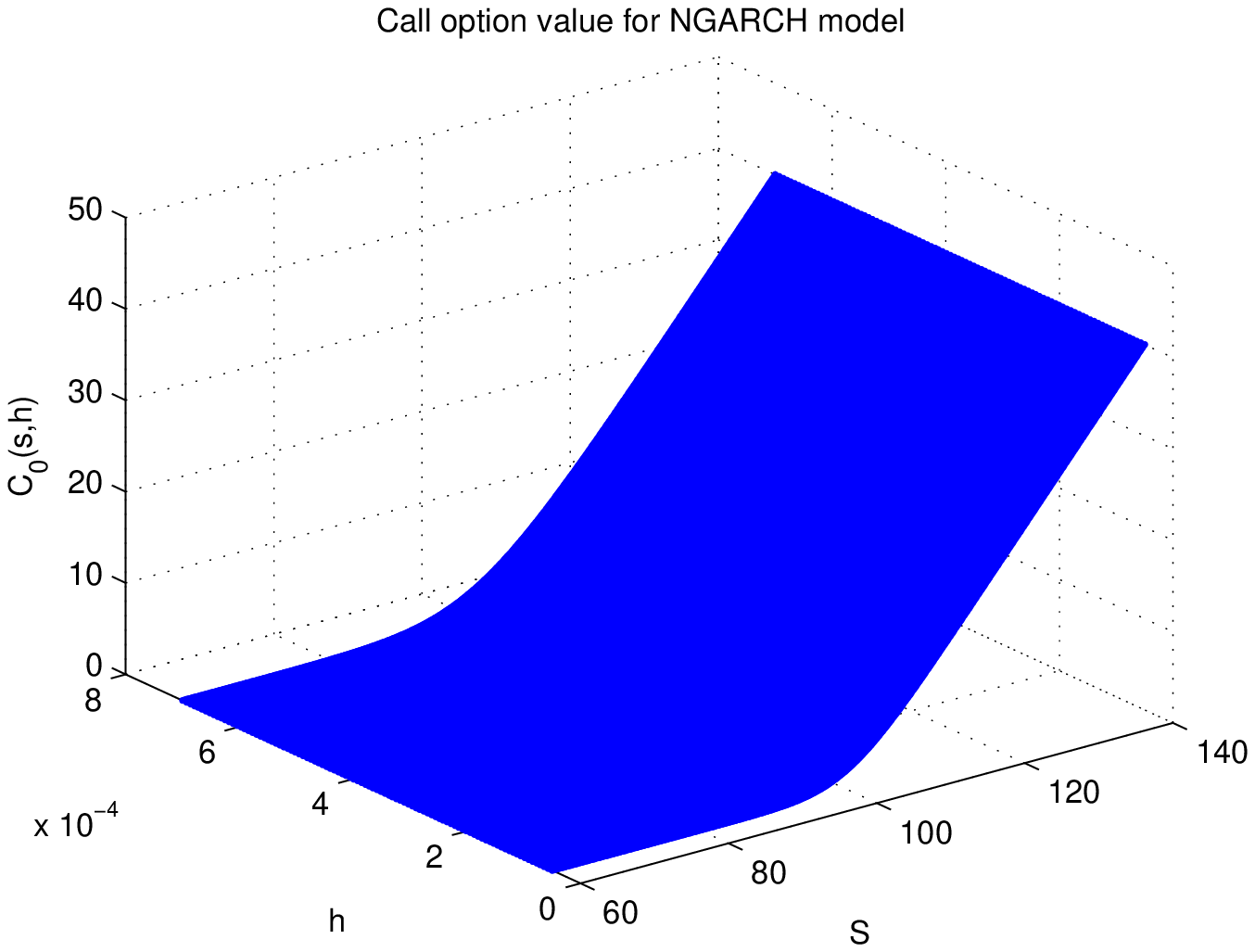}
\includegraphics[scale=0.375]{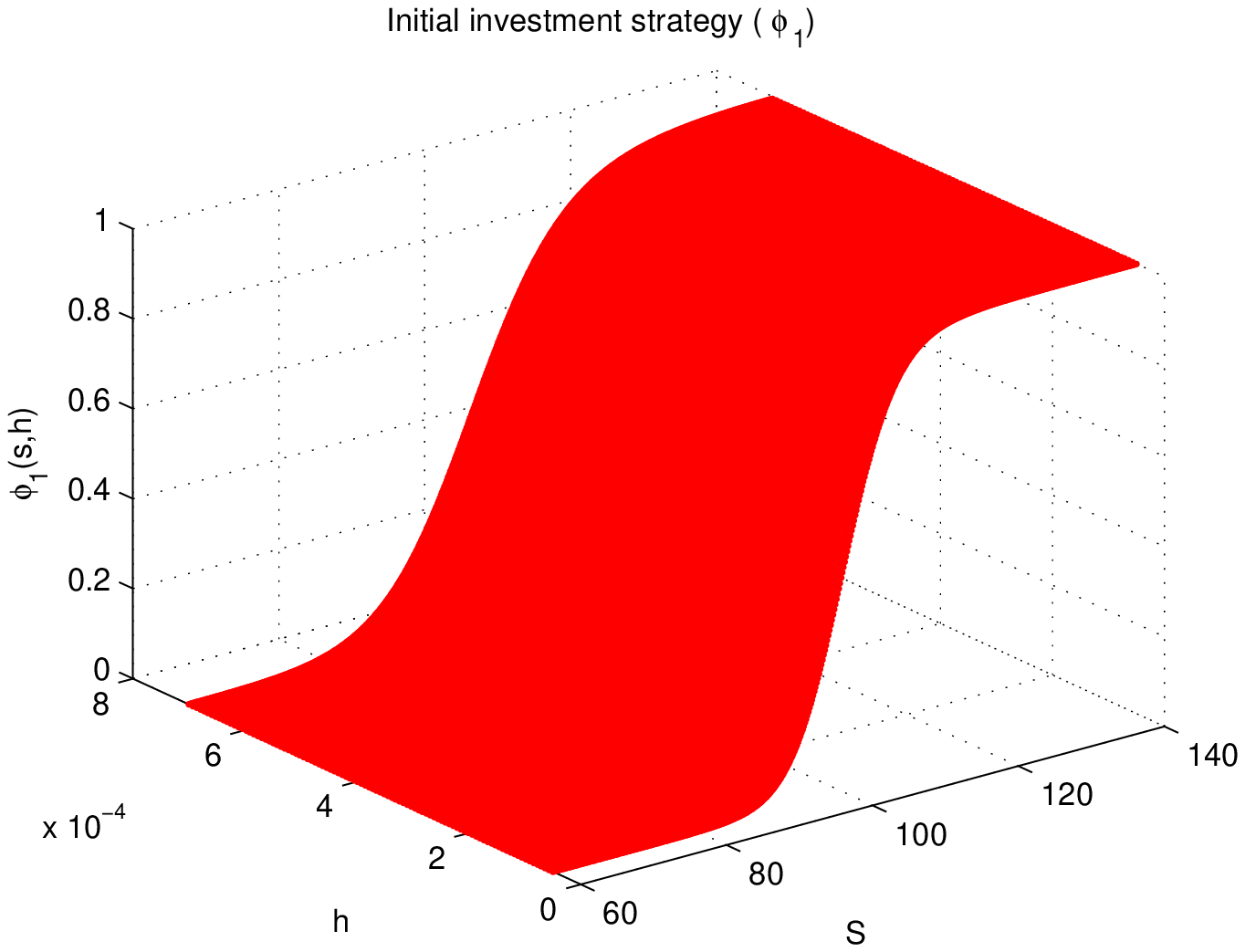}
\caption{Optimal hedging call option value $C_0$ and initial
investment strategy $\phi_1$ for the NGARCH model with $30$ periods.}\label{fig:NGarch}
\end{center}
\end{figure}

As in \citet{Duan:1995}, we consider the $NGARCH$ model where
$e^{\xi_k}-1 = r+ \lambda \sqrt{h_{k-1}} -\frac{1}{2}h_{k-1}
+\sqrt{h_{k-1}}\varepsilon_k$, and $ h_{k} = \alpha_0 + \alpha_1
h_{k-1}\varepsilon_k^2+ \beta_1 h_{k-1}$, with $\varepsilon_k\sim
N(0,1)$ and parameters $\alpha_0 = 1.524\times 10^{-5}$, $\alpha_1 =
0.1883$,       $\beta_1  = 0.7162$ and $\lambda = 7.452\times
10^{-3}$. Under the EMM, we have  $e^{\xi_k}-1  = r+
-\frac{1}{2}h_{k-1} +\sqrt{h_{k-1}}\varepsilon_k$ and $ h_{k} =
\alpha_0 + \alpha_1 h_{k-1}(\varepsilon_k-\lambda)^2+  \beta_1
h_{k-1}$. We price and hedge a call with strike $K=100$ and maturity
$30$ days using daily replication, using a grid of 500 points for
the asset on $[60,140]$, while the grid for the volatility consists
in 90 points of the interval $[.00005,.0007]$. The annual rate is
$0\%$. In what follows, B\&S hedging means delta hedging using the
B\&S formulas, while Duan's methods consists in picking his
suggested EMM and taking the delta of the option. The value of the
option and the initial number of asset are displayed in Figure
\ref{fig:NGarch}, while descriptive statistics of the 10000 hedging
errors are given in Table \ref{tab:NGarch} for the three hedging
methodologies, showing that the errors are more concentrated about 0
for the optimal hedging.

\begin{table}[h!]
{\footnotesize \caption{Statistics of hedging errors
(Payoff-Portfolio) for the NGARCH model.}\label{tab:NGarch}
\begin{center}
\begin{tabular}{l|r|r|r}
Stats      & Optimal        & Delta  &   Duan \\
\hline\hline
Average    &   -0.0159   &   -0.0954     &     0.0085         \\
Median     &   -0.1549    &  -0.2094       &     -0.1297          \\
Volatility &   0.8568     &   0.8951       &     0.9032           \\
Skewness   &   1.7205      &   2.7558        &     2.9171            \\
Kurtosis   &   10.5790       &   29.3709        &     29.9947            \\
Minimum    &   -1.9966     & -2.2302         &    -2.0789             \\
Maximum    &   9.9114       &  17.4873         &    17.7232             \\
VaR(99\%)  &   2.7698      &    2.8613       &   3.0831              \\
VaR(99.9\%)&   5.4893      &   6.3574        &    6.5070             \\
RMSE       &   0.8569     &  0.9001         &    0.9033            \\
\end{tabular}
\end{center}
}
\end{table}

\section{Conclusion}

In this paper we presented the optimal  solution for a discrete time
hedging portfolio. When the underlying  process is Markov or a
component of a Markov process, the optimal hedging strategy depends
on deterministic functions that can be approximated. We also find
explicit formulas for two interesting models.  Finally, numerical
simulations show that optimal hedging is preferable to delta
hedging.

\bibliography{all2005}

\begin{thebibliography}{}

\bibitem[Bouchaud and Potters, 2002]{Bouchaud/Potters:1999}
Bouchaud, J.-P. and Potters, M. (2002).
\newblock Back to basics: historical option pricing revisited.
\newblock {\em Philosophical Transactions: Mathematical, Physical \&
  Engineering Sciences}, 357(1758):2019 -- 2028.

\bibitem[Boyle and Emanuel, 1980]{Boyle/Emanuel:1980}
Boyle, P.~P. and Emanuel, D. (1980).
\newblock Discretely adjusted option hedges.
\newblock {\em Journal of Financial Economics}, 8:259--282.

\bibitem[{\v{C}}ern{\'{y}} and Kallsen, 2007]{Cerny/Kallsen:2007}
{\v{C}}ern{\'{y}}, A. and Kallsen, J. (2007).
\newblock On the structure of general mean-variance hedging strategies.
\newblock {\em Ann. Probab.}, 35(4):1479--1531.

\bibitem[Cornalba et~al., 2002]{Cornalba/Bouchaud/Potters:2002}
Cornalba, L., Bouchaud, J.-P., and Potters, M. (2002).
\newblock Option pricing and hedging with temporal correlations.
\newblock {\em Int. J. Theor. Appl. Finance}, 5(3):307--320.

\bibitem[Del~Moral et~al., 2006]{DelMoral/Remillard/Rubenthaler:2006}
Del~Moral, P., R\'emillard, B., and Rubenthaler, S. (2006).
\newblock {Monte Carlo approximations of American options}.
\newblock Technical report, GERAD.

\bibitem[Del~Moral et~al., 2012]{DelMoral/Remillard/Rubenthaler:2012}
Del~Moral, P., R\'emillard, B., and Rubenthaler, S. (2012).
\newblock {Monte Carlo Approximations of American Options that Preserve
  Monotonicity and Convexity}.
\newblock In {\em Numerical Methods in Finance}, pages 117--145. Springer.

\bibitem[Duan, 1995]{Duan:1995}
Duan, J.-C. (1995).
\newblock The {GARCH} option pricing model.
\newblock {\em Math. Finance}, 5(1):13--32.

\bibitem[Garcia and Renault, 1998]{Garcia/Renault:1998}
Garcia, R. and Renault, . (1998).
\newblock A note on hedging in {ARCH} and stochastic volatility option pricing
  models.
\newblock {\em Math. Finance}, 8(2):153--161.

\bibitem[Kat and Palaro, 2005]{Kat/Palaro:2005}
Kat, H.~M. and Palaro, H.~P. (2005).
\newblock Who needs hedge funds? {A} copula-based approach to hedge fund return
  replication.
\newblock Technical report, Cass Business School, City University.

\bibitem[Motoczy{\'n}ski, 2000]{Motoczynski:2000}
Motoczy{\'n}ski, M. (2000).
\newblock Multidimensional variance-optimal hedging in discrete-time model---a
  general approach.
\newblock {\em Math. Finance}, 10(2):243--257.
\newblock INFORMS Applied Probability Conference (Ulm, 1999).

\bibitem[Papageorgiou et~al., 2008]{Papageorgiou/Remillard/Hocquard:2008}
Papageorgiou, N., R\'emillard, B., and Hocquard, A. (2008).
\newblock Replicating the properties of hedge fund returns.
\newblock {\em Journal of Alternative Invesments}, 11:8--38.

\bibitem[R\'emillard et~al., 2010]{Remillard/Hocquard/Papageorgiou:2010}
R\'emillard, B., Hocquard, A., and Papageorgiou, N.~A. (2010).
\newblock {Option Pricing and Dynamic Discrete Time Hedging for
  Regime-Switching Geometric Random Walks Models}.
\newblock Technical report, SSRN Working Paper Series No. 1591146.

\bibitem[Schweizer, 1995]{Schweizer:1995a}
Schweizer, M. (1995).
\newblock Variance-optimal hedging in discrete time.
\newblock {\em Math. Oper. Res.}, 20(1):1--32.

\bibitem[Wilmott, 2006]{Wilmott:2006c}
Wilmott, P. (2006).
\newblock {\em Paul Wilmott on Quantitative Finance}, volume~3.
\newblock John Wiley \& Sons, second edition.

\end{thebibliography}
\bibliographystyle{apalike}

\appendix

\numberwithin{equation}{subsection}

\section{Proofs of the main results}\label{app:pfs}

\subsection{Proof of Lemma \ref{lem1}}\label{app:lem}

First, we will show that $\gamma_k \in (0,1]$ and $A_k$ is
invertible for all $k=1,\ldots, n$. By hypothesis,
$E(\gamma_{k+1}|\calf_{k-1})A_k - \mu_k\mu_k^\top $ is invertible
for all $k=1,\ldots, n$. In particular, it is true for $k=n$,
yielding that $\Sigma_n =A_n - \mu_n\mu_n^\top$ is invertible, which
is the conditional covariance matrix of $\Delta_n$ given
$\calf_{n-1}$. It then follows from Proposition \ref{prop:inv} that
$A_n$ is invertible. Without loss of generality, one may assume that
$A_n$ is diagonal. Otherwise, we diagonalize it in the form $A_n =
M_n B_n M_n^\top$, with $M_n,B_n$ $\calf_{n-1}$-measurable, $B_n$ is diagonal,
$M_n^\top M_n =I$ and set $\tilde \Delta_n = M_n^\top \Delta_n$.
Since $M_n M_n^\top =I$, it follows that $M_n$ is bounded, so
$\tilde \Delta_n$ is square integrable. Finally $B_n = E(\tilde
\Delta_n \tilde\Delta_n^\top|\calf_{n-1})$. $A_n$ being diagonal, it then follows that
$b_n^\top \Delta_n$ is square integrable and $\gamma_n = 1-b_n^\top \mu_n =
1-\mu_n^\top A_n^{-1}\mu_n$. It also follows from Proposition
\ref{prop:inv} that $\mu_n^\top A_n^{-1}\mu_n = \frac{\mu_n^\top
\Sigma_n^{-1}\mu_n}{1+\mu_n^\top \Sigma_n^{-1}\mu_n}$, so $\gamma_n
= \frac{1}{1+\mu_n^\top \Sigma_n^{-1}\mu_n} \in (0,1]$. As a result,
$\gamma_n \le E(\gamma_{n+1}\calf_n)=1$. The rest of the proof
follows easily by reverse
induction, using Proposition \ref{prop:inv} with the mean and covariance matrix of $\Delta_k$ under the probability
distribution $Q_k$, with $Q_k(O) = E(\I_O
\gamma_{k+1}|\calf_{k-1})/E(\gamma_{k+1}|\calf_{k-1})$, $O\in
\calf_{k}$, for $k=n-1, \ldots,1$. \qed

\subsection{Proof of Theorem \ref{thm1}}\label{app:thm}

Using the proof of Lemma \ref{lem1}, one can easily check that
$a_k$, $b_k$ and $\phi_k$ make sense and that $\phi_k^\top \Delta_k$
is square integrable. Next, it is easy to check that a necessary and
sufficient condition for $\left(V_0,\overrightarrow{\phi}\right)$ to
minimize
$E\left[\left\{G\left(V_0,\overrightarrow{\phi}\right)\right\}^2\right]$
is that $E\left\{G\left(V_0,\overrightarrow{\phi}\right)\right\}=0$
and
$E\left\{G\left(V_0,\overrightarrow{\phi}\right)\Delta_k|\calf_{k-1}\right\}=0$
for all $k=1,\ldots,n$. The necessity comes from the fact that for
any event $O\in \calf_{k-1}$, one must have $
0=\left.\frac{d}{d\epsilon}\right|_{\epsilon=0}
E\left[\left\{G\left(V_0,\overrightarrow{\phi}\right) -\epsilon \I_O
\Delta_k\right\}^2 \right]=
-2E\left\{G\left(V_0,\overrightarrow{\phi}\right)\Delta_k
\I_O\right\}$, which is equivalent to the condition
$E\left\{G\left(V_0,\overrightarrow{\phi}\right)\Delta_k|\calf_{k-1}\right\}=0$,
while the condition
$E\left\{G\left(V_0,\overrightarrow{\phi}\right)\right\}=0$ comes
from the fact that for any $\theta$, one must have
$$
0=\left.\frac{d}{d\epsilon}\right|_{\epsilon=0}
E\left[\left\{G\left(V_0+\epsilon
\theta,\overrightarrow{\phi}\right) \right\}^2 \right]=
-2E\left\{G\left(V_0,\overrightarrow{\phi}\right)\right\}.
$$
To see that the conditions are sufficient, it suffices to check that
$$
E\left[\left\{G\left(V_0+\theta_0,\overrightarrow{\phi+\psi}\right)\right\}^2 \right] =
E\left[\left\{G\left(V_0,\overrightarrow{\phi}\right)\right\}^2 \right]+ E\left\{\left(\theta_0+\sum_{k=1}^n \psi_k^\top \Delta_k \right)^2\right\}.
$$
The proof that $\overrightarrow{\phi}$ is the solution is based on
the following equation, which can be easily proven by induction.

\begin{equation}\label{eq:V}
E(V_n|\calf_k) = V_kE(P_{k+1}|\calf_k) + E\left\{\beta_n
C(1-P_{k+1})|\calf_k \right\}, \quad k=1,\ldots,n.
\end{equation}
%
%
%

To complete the proof of theorem, note that  from \eqref{eq:V},
\begin{equation}\label{eq:G}
E\left\{ G\left(V_0,\overrightarrow{\phi}\right)|\calf_k\right\} =
E(\beta_n C P_{k+1}|\calf_k)-V_k E(P_{k+1}|\calf_k), \quad
k=0,\ldots, n.
\end{equation}
Using \eqref{eq:G}, one has $ E\left\{
G(\left(V_0,\overrightarrow{\phi}\right)\Delta_k|\calf_k\right\} =
E(\beta_n C \Delta_k P_{k+1}|\calf_k)- E(V_k\Delta_k
P_{k+1}|\calf_k)$, so $ E\left\{
G(\left(V_0,\overrightarrow{\phi}\right)\Delta_k|\calf_{k-1}\right\}
= E(\beta_n C \Delta_k P_{k+1}|\calf_{k-1})- E(V_k\Delta_k
P_{k+1}|\calf_{k-1})=  A_k (a_k- V_{k-1}b_k -\phi_k) =0$. Hence
$ E\left\{
G(\left(V_0,\overrightarrow{\phi}\right)\right\} = E(\beta_n C
P_{1})-V_0 E(P_{1})=0$. \qed
%

\subsection{Proof of Proposition \ref{prop:ck}}\label{app:res2}

The result  is obviously true for $k=n+1$. Suppose it is true for
$k+1$. For $i$ given, set $\pi_j = Q_{ij}\gamma_{k+1}(j)/D$, where
$D =\sum_{j=1}^l Q_{ij}\gamma_{k+1}(j)$. By hypothesis, $\pi_1,
\ldots, \pi_l$ are probabilities adding to $1$, so if $X \sim \mathbb{P}_j$ with
probability $\pi_j$, then $ \gamma_k(i) = D\left(1- \mu^\top
B^{-1}\mu \right)$, where $\mu = E(X)$ and $B  = E\left(X
X^\top\right)$. Let $\Sigma$ be the covariance matrix of $X$. It is
non singular since the covariance of $X$ under $\mathbb{P}_j$ is non
singular. It then follows from Proposition \ref{prop:inv} that $1-
\mu^\top B^{-1}\mu =\frac{1}{1+\mu^\top \Sigma^{-1}\mu}>0$. Since
$D>0$ by hypothesis, one may conclude that $\gamma_k(i)>0$. As a
by-product we get that $\gamma_k(i) \le 1$ if $\gamma_{k+1}(j)\le 1$
for all $j=1,\ldots$. Since that  is true for $\gamma_{n+1}\equiv
1$, one may conclude that for all $k=1,\ldots, n$, $\gamma_k(i)\le
1$. The rest of the proof is easy. \qed

\section{Auxiliary results}

\begin{prop}\label{prop:inv}
Suppose $A = \Sigma+ bb^\top$ where $\Sigma$ is symmetric and
invertible. Then $A$ is invertible, and $ A^{-1} = \Sigma^{-1} -
\frac{\Sigma^{-1} b b^\top \Sigma^{-1}}{1+ b^\top \Sigma^{-1} b}$.
Moreover, $1- b^\top A^{-1}b = \frac{1}{1+ b^\top \Sigma^{-1} b}
>0$.
\end{prop}

Proof: Since $A \left(\Sigma^{-1} - \frac{\Sigma^{-1} b b^\top
\Sigma^{-1}}{1+ b^\top \Sigma^{-1} b} \right)=I$,  $A$ is invertible
and $A^{-1} = \Sigma^{-1} - \frac{\Sigma^{-1} b b^\top
\Sigma^{-1}}{1+ b^\top \Sigma^{-1} b}$. Setting $c = b^\top
\Sigma^{-1} b$, one gets $ 1- b^\top A^{-1}b = 1 -c +
\frac{c^2}{1+c} = \frac{1}{1+ c} >0$.
 \qed

\end{document}